\documentclass[5p,sort&compress,number]{elsarticle}

\usepackage{amssymb}
\journal{Nuclear Instruments and Methods}

\begin{document}

\begin{frontmatter}



\title{High Energy Neutrino Telescopes in the Northern Hemisphere}


\author{Juan Jos\'e Hern\'andez--Rey}

\address{IFIC - Instituto de F\'{\i}sica Corpuscular,  
Universitat de Val\`encia--CSIC, E-46100 Valencia, Spain}

\ead{Juan.J.Hernandez@ific.uv.es}

\begin{abstract}
We review the status and results of the high energy neutrino
telescopes in the Northern Hemisphere, namely ANTARES and Baikal
(NT200+). After a brief introduction to Neutrino Astronomy, we
describe these telescopes in their past and present configurations and
report briefly on the results obtained in several areas, such as the
search for high energy cosmic neutrino diffuse fluxes and point
sources, the indirect search for dark matter, the multimessenger
studies and the search for exotic particles, such as monopoles and
nuclearites.

\end{abstract}

\begin{keyword}
Neutrino Telescopes \sep Neutrino Astronomy \sep High Energy Astrophysics

\PACS 95.55.Vj \sep 95.85Ry \sep 96.50 Pw


\end{keyword}

\end{frontmatter}

%%JJHR:
%\linenumbers

\section{Introduction}
\label{intro}

The detection of high energy cosmic neutrinos can help solve the
problem of the origin of high energy cosmic rays and be a new tool to
elucidate the mechanisms of hadronic acceleration in astrophysical
objects.  In the low energy domain (few MeV to several GeV) the
observation of extraterrestrial and atmospheric neutrinos gave rise to
the discovery of neutrino oscillations and to one of the most direct
experimental tests of our models of supernova explosions. In the high
energy regime (several GeV to EeV), neutrinos have several advantages
as cosmic messengers and can provide information on the particle
acceleration mechanisms in the Universe. Experimental methods to
detect them exist and have been technologically proven. The major
challenge in the field of Neutrino Astronomy is at present to reach a
sensitivity high enough to detect the first cosmic neutrino sources.

Let us briefly summarize the advantages of neutrinos as cosmic
messengers. They are neutral particles, therefore they are not
deflected by magnetic fields and point back to their sources. They are
weakly interacting and thus can escape from very dense astrophysical
objects and travel long distances without being absorbed by matter or
background radiation.  Moreover, in cosmic sites where hadrons are
accelerated, it is likely that neutrinos are generated in the decay of
charged pions produced in the interaction of those hadrons with the
surrounding matter or radiation, being therefore a smoking gun of
hadronic acceleration mechanisms.

The observation of neutrinos in a Cherenkov neutrino telescope is
based on the detection of the muons produced by the neutrino charged
current interactions with the matter surrounding the telescope by
means of the Cherenkov light induced by the muons when crossing the
detector medium, natural ice or water. Cascades produced in charged or
neutral current interactions of neutrinos inside or nearby the
detector can also be detected.

A typical neutrino telescope consists of a three dimensional array of
light sensors, photomultipliers (PMTs), that record the position and
time of the emitted Cherenkov photons, enabling the reconstruction of
the muon track or the cascade.  To avoid the huge background of muons
produced by cosmic ray showers in the atmosphere, the telescopes look
at the other side of the Earth, i.e. they use it as a shield against
the muons produced in normal atmospheric showers.  The increase in the
range of muons in the rock at high energies (from kilometres to
several kilometres) together with the increase of the neutrino cross
section gives rise to an approximately exponential increase of the
effective areas of these devices in the GeV to PeV energy range. Above
a few TeV the telescopes can determine the direction of the incoming
neutrinos with angular resolutions better than 1$^\circ$, hence the
name ``telescope''.  At energies above the PeV, the Earth becomes
opaque to neutrinos, but the atmospheric muon flux decreases
dramatically so that the neutrino telescopes can look for downgoing
neutrinos.  Other neutrino flavours can be observed through the
detection of hadronic or electromagnetic showers or, in the case of
tau neutrinos, via the observation of its interaction and the
subsequent decay of the produced tau lepton.

The first attempt to build a neutrino telescope in natural water,
namely the DUMAND project, dates back to the 60's~\cite{dumand}.
DUMAND paved the way for subsequent projects.  NT200 in Lake Baikal
and then ANTARES in the Mediterranean Sea benefited from the
experience of DUMAND. We will cover in this article these two last
experiments which are the first underwater neutrino telescopes ever
built.

The first efforts to install a neutrino telescope in Lake Baikal
started in the 80's~\cite{baifirst}.  After some site tests, the first
single string arrays were operated between 1984 and 1990. Already
since 1987 the construction of a telescope with 200~PMTs was
envisaged. Between 1993 and 1994, the so-called NT-36 version of this
telescope with 36 PMTs was operated. Since then the detector has been
growing gradually: NT-72 (1995-1996), NT-96 (1996-1997), NT-144
(1997-1998) and NT-200 (since 1998). In 2005 three outer strings were
added to form the so-called NT-200+.

\begin{figure}[bh]
\begin{center}
\includegraphics[scale=0.38]{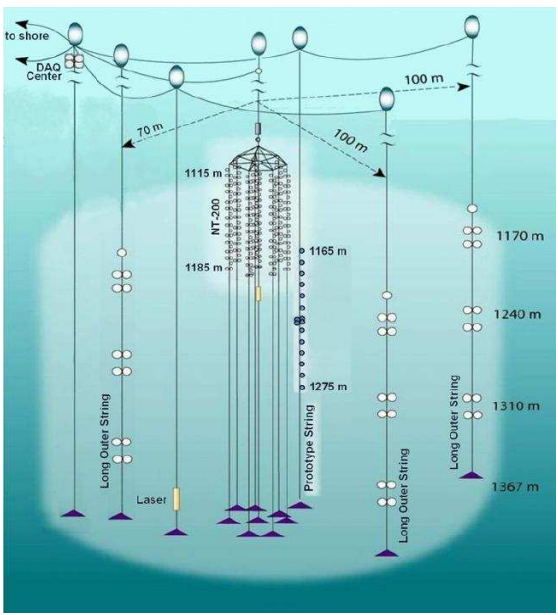}
\end{center}
\caption{\label{nt200} Schematic view of the Baikal Neutrino Telescope}
\end{figure}

The Baikal neutrino telescope NT-200 is located in Lake Baikal at a
latitude of around 52$^{\circ}$ North~\cite{bainim}.  The detector is
around 3.6~km from the shore and at a depth between 1115~m and
1185~m. The NT200 configuration is composed of 8 strings that are held
by an umbrella shaped mechanical structure (see Fig.~\ref{nt200}).
Each string has 24 optical modules (OMs) arranged in pairs adding up
to a total of 192 OMs. Each OM contains a 37-cm diameter QUASAR
photomultiplier~\cite{quasar} specifically designed for this
detector. The two photomultipliers (PMTs) of a pair are operated in
coincidence in order to supress background from bioluminiscence and
PMT noise. The upgraded version NT200+ includes three new lines that
surround the old NT200 detector and are located at 100~m from its
centre, thereby increasing its sensitivity by a factor four for very
high energy cosmic neutrinos.

\begin{figure}[t]
\begin{center}
\includegraphics[scale=0.42]{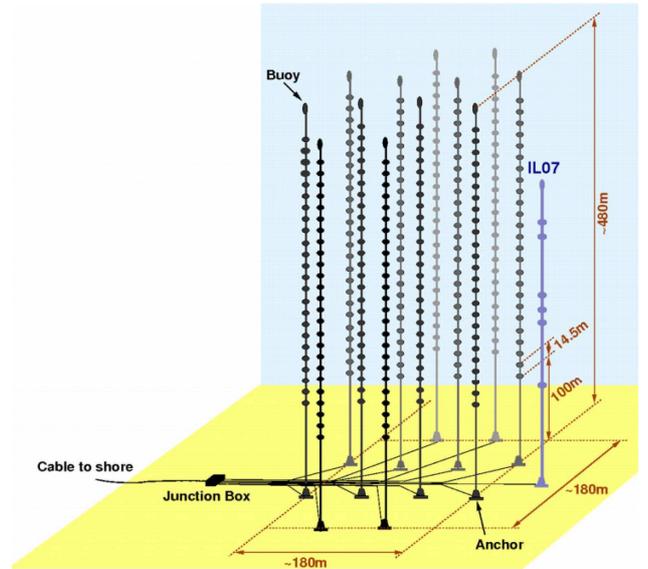}
\end{center}
\caption{\label{antares} Schematic view of the Antares Neutrino Telescope}
\end{figure}

The ANTARES neutrino telescope is located 40~km offshore from Toulon
at 2475~m depth at a latitude around 43$^{\circ}$
North~\cite{antaresdet}. It consists of 12 mooring lines anchored to
the sea bed and held taut by means of buoys (see Fig.~\ref{antares}).

\noindent

Each line contains 25 storeys.  The lowest storey is 100~m above the
sea bed and the vertical distance between consecutive storeys is
14.5~m.  The total line length is 480~m.  Each storey has a triplet of
OMs and an electronics module. The OMs contain a 10-inch
photomultiplier looking 45$^\circ$ downwards~\cite{om, pmt}.  In
addition, several optical beacons~\cite{obs} are distributed
throughout the lines for calibration purposes~\cite{timecalib}. The
horizontal separation between lines is between 60 and 80~m. Each line
is connected to a junction box by means of interlink cables and the
junction box is connected to the shore by the main electro-optical
cable.

The ANTARES initiative started in 1998 and after a period of site
evaluation, detector design, tests and construction, the first line
was deployed in 2006. The detector was operated with 5-lines during
several months in 2007 and was fully deployed in 2008. Taking
advantage of the possibility of detector maintenance offered by water,
the ANTARES collaboration has recovered and repaired some of the lines
and fixed problems in some of the detector's interlink cables.

\noindent

\section{Search for a diffuse flux of cosmic neutrinos}
The term diffuse flux refers to the search of cosmic neutrinos without
requiring precise directional information.  An excess over the
expected atmospheric neutrino background is looked for and if none is
found limits are customarily set on the normalization of signal fluxes
with energy spectra of the type $E^{-2}$.

The Baikal collaboration has performed several searches for diffuse
fluxes of cosmic neutrinos~\cite{baidiffuse1,baidiffuse2,
  baidiffuse3}.  They looked for cascades produced both in charged and
neutral current interactions of neutrinos in the medium surrounding
the detector. Initial cuts were applied to the energy of the
reconstructed cascades in order to select neutrino events. The number
of upward going cascades detected agrees with those expected from
background.  A cut on energy of 10~TeV and 130~TeV for upgoing and
downgoing cascades, respectively, was then introduced to select the
final neutrino signal.  No events were observed for an expected
background of around 2 events.  From this lack of signal an upper
limit of E$^{-2}\Phi < 2 \times 10^{-7}$ GeV cm$^{-2}$ s$^{-1}$
sr$^{-1}$ was set for the flux of all flavours of neutrinos of cosmic
origin in the energy interval 20~TeV $<$ E$_{\nu}$ $<$ 20~PeV. As
usual, a flavour ratio $\nu_{e}$:$\nu_{\mu}$:$\nu_{\tau}$ = 1:1:1 was
assumed.

Using the data collected by ANTARES during the period from December
2007 to December 2009, corresponding to a total live time of 334 days
with different detector configurations (9, 10 and 12 lines), a search
for a diffuse flux of astrophysical muon neutrinos was
performed~\cite{antdiffuse}. In addition to the cuts on the quality of
the reconstructed track and on the number of hits, an energy cut was
also applied. A novel technique based on the repetition rate of
photoelectrons on a given PMT averaged over all the PMTs was used.
This variable is a good proxy of the energy of the track and is well
described by the Monte Carlo simulation. After unblinding of the data,
the number of events above the optimized cut in repetition rate was
found to agree with background expectations.  From the compatibility
of the observed number of events with the expected background and
assuming an E$^{-2}$ flux spectrum for the signal, a 90\% C.L. upper
limit on the $\nu_{\mu }+\bar{\nu}_{\mu }$ diffuse flux of E$^2 \,
\Phi <$ 5.3 $\times$10$^{-8}$ GeV cm$^{-2}$ s$^{-1}$ in the energy
range 20~TeV to 2.5~PeV was obtained.

\begin{figure}[hbt]
\begin{center}
\includegraphics[scale=0.85]{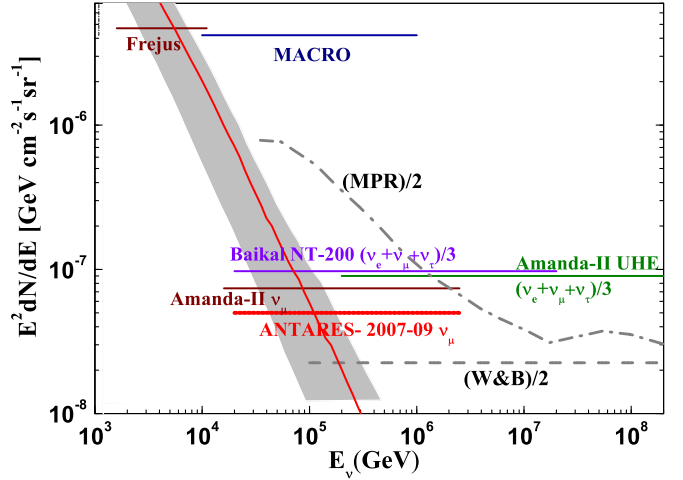}
\end{center}
\caption{90\% C.L. upper limits for an E$^{-2}$ high energy cosmic
  neutrino diffuse flux from several experiments(see text for
  explanations). }
\label{antdiffuse}
\end{figure}

The 90\% C.L. upper limits on a diffuse flux of muon neutrinos (plus
antineutrinos) from several experiments~\cite{otherdiffuse} are given
in Fig.~\ref{antdiffuse}.  The original limits given by BAIKAL NT-200
and Amanda-II UHE are for all flavours and are divided by 3 in this
plot for the sake of comparison.  The recent limit on the diffuse flux
of astrophysical $\nu_{\mu}$ from the IceCube experiment, E$^2 \, \Phi
<$ 8.9 $\times$10$^{-9}$ GeV cm$^{-2}$ s$^{-1}$~\cite{icediffuse}, is
not shown in this figure. The grey band represents the expected
variation of the atmospheric $\nu_{\mu}$ flux: the minimum is the
Bartol flux from the vertical direction and the maximum is the
Bartol+RQPM flux from the horizontal direction. The central line is
averaged over all directions. The phenomenological upper bounds of
W\&B and MPR~\cite{WB} are also given, dividing by 2 the original
values in order to take into account neutrino oscillations.

\section{Search for point-like sources}
\label{pointsources}

A search for point sources was performed by ANTARES using the data
taken from 2007 to 2010. After the selection of data runs requiring
that most of the detector was operating and that the optical
background from bioluminiscence was low, the final data sample
amounted to a total of 813 live days.  Only events with upgoing muons
were kept for further analysis, requiring in addition that the
corresponding track had a good reconstruction quality and an estimated
angular error lower than 1$^{\circ}$. The cut in quality was chosen so
as to optimize the discovery potential. A total of 3058 events were
selected. According to Monte Carlo simulations around 15\% of them
were atmospheric muons wrongly reconstructed as upgoing tracks.

Clusters of events with a large enough significance above that
expected from background fluctuations were looked for with a likehood
ratio method.  The likehood used the distribution in declination of
the atmospheric background obtained by scrambling the data in right
ascension and an angular resolution of (0.46$\pm$0.15)$^{\circ}$, as
given by Monte Carlo simulation.  The full sky was searched for
possible sources and then a list of 51 pre-selected directions in the
sky corresponding to possible astrophysical neutrino sources were
scrutinized. No significant excess was found in either case.  An
alternative search method~\cite{em} was used as a cross-check
obtaining similar results.

\begin{figure}[htb]
\begin{center}
\includegraphics[scale=0.45]{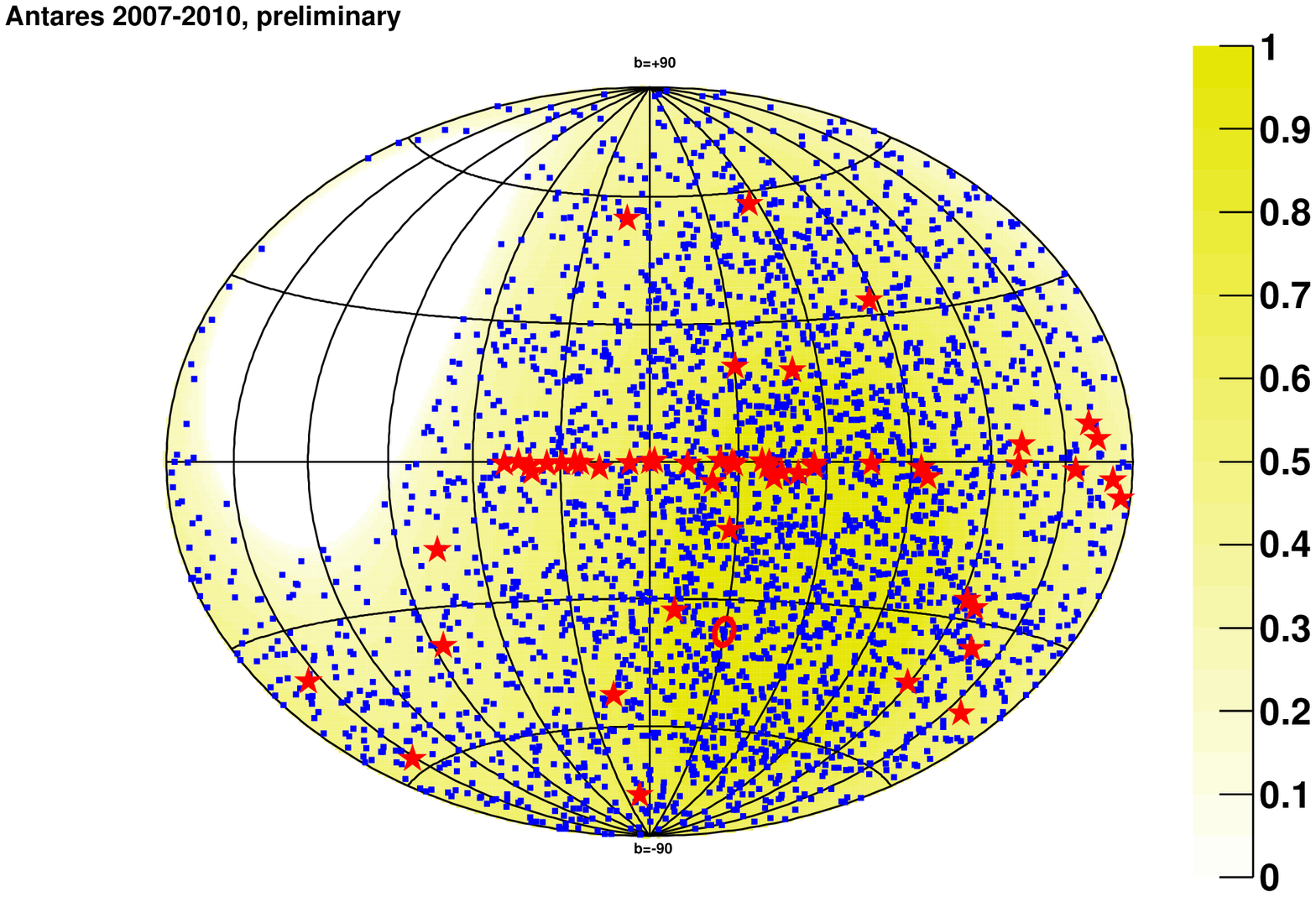}
\end{center}
\caption{\label{skymap} Skymap in Galactic coordinates. The grade in
  the hue of the (yellow) background indicates the visibility of the
  corresponding region according to the scale on the right (white:
  0\%; darkest yellow: 100\%). Blue dots: position in the sky of the
  3058 selected neutrinos candidates. Red stars: position of the 51
  pre-selected sources. Red ellipse: the most significant cluster of
  events.}
\end{figure}

In Fig.~\ref{skymap} the direction in Galactic coordinates of all the
selected tracks are shown as (blue) dots. The hue of the yellow
background of the figure indicates the percentage of visibility of the
corresponding region of the sky, white corresponds to no visibility
and dark yellow to 100\%.

\begin{figure}[htb]
\begin{center}
\includegraphics[scale=0.45]{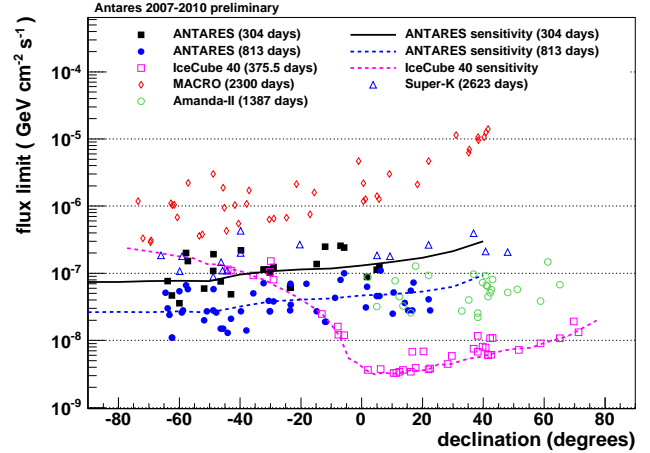}
\end{center}
\caption{\label{pslimits} 90\% C.L. upper limits for a neutrino flux
  with an E$^{-2}$ spectrum for 51 candidates sources (blue points)
  and the corresponding sensitivity (dashed blue curve). Results from
  the MACRO, Amanda~II, Super Kamiokande and IceCube
  telescopes~\cite{psothers} are also shown.  }
\end{figure}

The most significant cluster in the full-sky search was found at
$\alpha = -46.5^{\circ}$ and $\delta= -65.0^{\circ}$ and is indicated
in Fig.~\ref{skymap} by a (red) ellipse. The post-trial $p$-value for
this cluster was 2.5\%, a value not significant enough to claim a
signal.

The (red) stars in the figure correspond to the sky position of the 51
pre-selected sources for which the dedicated candidate search was
carried out. The most significant source of the predefined list (HESS
J1023-575) was fully compatible with a background fluctuation
($p$=41\%). The corresponding limits for neutrino sources emitting
with an $E^{-2}$ energy spectrum are given in Fig.~\ref{pslimits}.
This limit is 2.5 times better than the one previously
published~\cite{psprevious}. Limits from other experiments are also
given~\cite{psothers}. As can be seen, these results are at present
the most stringent for the Southern Sky, except for the case of the
IceCube detector for which in this hemisphere very high energy
neutrinos can be looked for (E$>$ 1~PeV).  Note that, even though
neutrino sources in the Galaxy with a UHE component are not
discarded~\cite{pevatrons}, for the more plausible Galactic neutrino
sources (e.g. young SNRs) most of the neutrino signal is expected to
lie below a few hundred TeV~\cite{galneut, galneutb}.

\section{Multimessenger searches}

The search of neutrinos in coincidence with other messengers has
several advantages. Sources already known to have high-energy
emission, e.g. gamma-rays, can be investigated, increasing the chance
to observe sites of hadronic acceleration. In addition, the
restriction of the search to limited time windows and sky directions
highly reduces the atmospheric neutrino background and therefore
increases the sensitivity to possible signals, so that a handful of
events can be enough to claim a signal. In the case of neutrino events
coincident with gravitational waves, the same astrophysical phenomena
are expected to produce both types of signals.  We give below a couple
of examples of the multimessenger program, which is too broad to be
fully reported here.

A selection of flares from blazars observed by the LAT detector of the
Fermi satellite during 2008 was carried out and the data taken by
ANTARES in the same period was investigated for neutrino coincidences
with the flaring period of the blazars~\cite{blazars}. The selected
blazars are shown in Table~\ref{tab:blazars} together with the number
of events required to claim a 5$\sigma$ signal. Only one event
--during a flare of 3C279-- was detected. The post-trial $p$-value of
such a coincidence is 10\%, compatible with a background
fluctuation. The 90\% C.L. limits on the neutrino fluence from these
blazars are given in Table~\ref{tab:blazars} for the 100~GeV to 1~PeV
region and assuming an $E^{-2}$ spectrum.

\begin{table}[ht]
      \begin{center}
      \begin{tabular}{|l|c|c|}
      \hline 
Source & N($5\sigma$) & Fluence\\ \hline \hline 
PKS0208-512 & 4.5 & 2.8  \\ \hline 
AO0235+164  & 4.3 & 18.7 \\ \hline
PKS1510-089 & 3.8 & 2.8  \\ \hline 
3C273       & 2.5 & 1.1  \\ \hline
3C279       & 5.0 & 8.2  \\ \hline 
3C454.3     & 4.4 & 23.5 \\ \hline
OJ287       & 3.9 & 3.4  \\ \hline 
PKS0454-234 & 3.3 & 2.9  \\ \hline
WComae      & 3.8 & 3.6  \\ \hline 
PKS2155-304 & 3.7 & 1.6  \\ \hline
      \end{tabular}
\caption{List of blazars for which neutrinos were looked for in
  coincidence with their flares. {\it N(5$\sigma$)} is the average
  number of events required for a 5$\sigma$ discovery (50\%
  probability) and {\it Fluence} is the upper limit (90\% C.L.) on the
  neutrino fluence in GeV$\cdot$cm$^{-2}$.}
      \label{tab:blazars}
      \end{center}     
\end{table}

\begin{figure}[t]
\begin{center}
\includegraphics[scale=0.60]{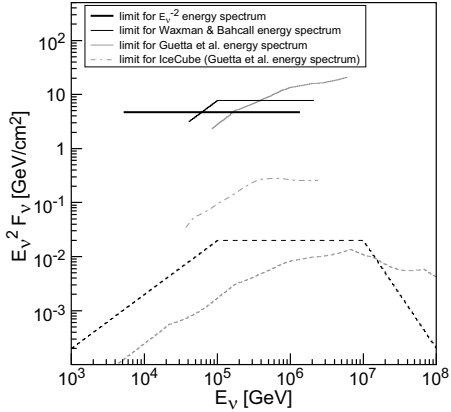}
\end{center}
\caption{\label{grbs} 90\% C.L. limits on the total muon neutrino
  fluence from the selected 40~GRBs as a function of the neutrino
  energy for three different energy spectra: E$^{-2}$ (thick solid
  line), Waxman and Bahcall GRB model~\cite{grbwb} and Guetta et
  al.~\cite{grbguet}.  The black dashed line is the expected neutrino
  fluence for 40 GRBs with the Waxman and Bahcall energy spectrum and
  the grey dashed line is the sum of the 40 expected individual GRB
  fluences according to Guetta et al. The total prompt emission
  duration of the 40 GRBs is 2114~s. The grey dash--dotted line is the
  90\% C.L. limit set by IceCube using 117 GRBs~\cite{grbice}. }

\end{figure}

Several models predict the production of high energy neutrinos during
gamma-ray bursts. As in the previous analysis, restricting the search
to a short time window sizeably reduces the atmospheric background so
that only a few events would be enough to claim a discovery. Using the
2007 ANTARES data, a search for neutrinos coming from 40 GRBs events
was performed. No neutrino event was found in the corresponding time
windows and within the defined search cone around each source.  The
limits obtained from this lack of signal are shown in Fig.~\ref{grbs},
where the 90\% C.L. limits on the total fluence of the 40 GRBs are
shown as a function of the neutrino energy for three different energy
spectra.

The Baikal collaboration has also performed a search for neutrinos
associated to 303 GRBs alerts provided by the BATSE detector from 1998
to 2000~\cite{grbbai}. From the absence of neutrino events in
coincidence with the GRBs a 90\% C.L. limit on the neutrino flux of
E$^{-2} \Phi_{\nu} <$ 1.1 $\times$ 10$^{-6}$ GeV cm$^{-2}$ s$^{-1}$
was set for a Waxman and Bahcall--type energy spectrum~\cite{grbwb}.

\section{Indirect search for dark matter}
If dark matter is made up of weakly interacting massive particles
(WIMPs), some of these will slow down by elastic scattering and end up
gravitationally trapped in heavy astrophysical objects such as the
centre of the Galaxy, the Sun or the Earth. They will then
self-annihilate and high energy neutrinos will be emitted in the decay
chain of their products.

\begin{figure}[htb]
\begin{center}
\includegraphics[scale=0.43]{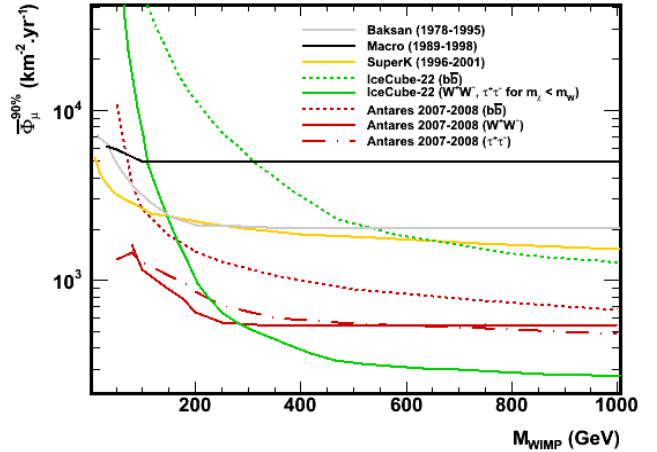}
\end{center}
\caption{\label{wimps} ANTARES 90\% sensitivity on the muon flux as a
  function of the WIMP mass using the 2007-2008 data.  Also shown are
  the 90\% C.L. limits on the same flux set by different experiments:
  with Baksan~\cite{dmbak}, Macro~\cite{dmmac},
  SuperKamiokande~\cite{dmsk}, and IceCube-22~\cite{dmic} for the $b
  \bar{b}$ and $W^+ W^-$ channels).  }
\end{figure}

ANTARES is analysing the data taken during 2007 and 2008, a total live
time of around 300 days, looking for high energy neutrinos coming from
the Sun.  The analysis is based on the optimization of the cuts based
on the reconstruction quality of the muon tracks and the size of the
half-cone angle around the Sun direction to select neutrino
candidates.  The sensitivity, i.e. the expected average 90\%
C.L. upper limit, for the muon flux coming from the Sun is shown in
Fig.~\ref{wimps} for the CMSSM framework. As expected the more
stringent limits will come from the hard channels, i.e those that
produce $W^+ W^-$ or $\tau^+ \tau^-$ in the annihilation process.
 
Using the data recorded during the period 1998--2002, a total of 1007
live days, the Baikal collaboration was able to set a 90\% C.L. upper
limit of $\Phi < 3 \times 10^3$ km$^{-2}$ yr$^{-1}$ on an excess in
the muon flux coming from the Sun for neutralino masses larger than
100~GeV~\cite{dmbai}.  Similarly using a total of 1038 live days, the
corresponding 90 \% C.L. upper limit for a flux coming from the
Earth's core was found to be $\Phi < 1.2 \times 10^3$ km$^{-2}$
yr$^{-1}$ for neutralino masses greater than 100~GeV.

\section{Search for exotic particles}

The existence of monopoles has been put forward in the context of
several theories. To date there is no clear evidence of their
existence and several limits have been set on the flux of monopoles
crossing the Earth.

Relativistic monopoles with masses above 10$^7$ GeV can cross the
Earth and leave a conspicuous signal in neutrino telescopes.  Magnetic
charges crossing water at a speed larger than their Cherenkov
threshold ($\beta > 0.74$ in water) would produce a huge amount of
light. For one unit of magnetic charge this radiation would be 8550
times larger than that of a muon. Moreover, even below the threshold,
for $\beta > 0.52$, the high energetic ionization electrons
($\delta$-rays) produced by the monopole would also radiate a large
amount of light.

\begin{figure}[htb]
\begin{center}
\includegraphics[scale=0.47]{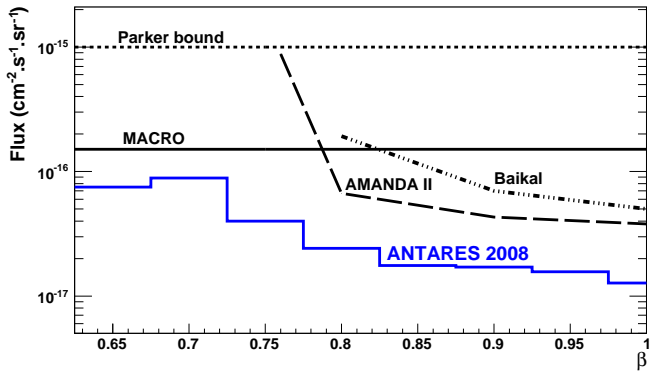}
\end{center}
\caption{\label{monopoles} 90\% C.L. upper limit on a flux of magnetic
  monopoles set by the ANTARES~\cite{antmonopoles}, 
Amanda~II~\cite{amanmonopoles}, Baikal~\cite{baimonopoles} and 
MACRO~\cite{macromonopoles} are also shown. }
\end{figure}
\noindent

The Baikal collaboration used the data taken between April 1998 and
February 2003, which includes the NT36, NT96 and NT200 configurations,
to perform a search for relativistic monopoles~\cite{baimonopoles}.
Events were selected on the basis of a high number of hits and good
reconstruction quality as determined by a $\chi^2$ test. Only upgoing
tracks (zenith angle greater than 100$^{\circ}$) were kept. A cut on
the radial distance depending on the exact detector configuration was
also applied. No candidate was found when simulations indicated that
around 4 background events were expected.  The 90 \% C.L. limits on
the flux obtained from this negative result are shown in
Fig.~\ref{monopoles}.

Using the data taken during 2007 and 2008, ANTARES performed a search
for magnetic monopoles based also on the quality of the track and the
number of hits as well as on the track reconstructed velocity,
$\beta$.  A special reconstruction was performed in which the $\beta$
of the particle was a free parameter and the $\chi^2$ values for the
hypotheses of $\beta$ equal or different from one were compared. The
selection criteria were optimized for discovery in eight velocity
intervals in the region 0.625 $\le \beta \le$ 0.995. Only one
candidate was found, compatible with the total expected background. In
Fig.~\ref{monopoles} the 90\% C.L. upper limit on the flux of upgoing
monopoles obtained is shown~\cite{antmonopoles}. As can be seen, this
limit is more stringent than the previous existing
limits~\cite{amanmonopoles, macromonopoles}.

A search for nuclearites, massive aggregates of up, down and strange
quarks, has also been performed by ANTARES. Nuclearites would produce
in water a thermal shock wave emitting a large amount of radiation at
visible wavelengths. No clear indication of nuclearites was observed
using the 2007-2008 data sample and a 90\% C.L. upper limit of
10$^{-16}$ cm$^{-2}$ sr$^{-1}$ s$^{-1}$ for a flux of nuclearites with
masses between 10$^{-14}$ and 10$^{-17}$ GeV was established.

\section{Summary}
The underwater neutrino telescopes ANTARES and NT-200 not only have
shown the technical and scientific feasibility of this sort of devices
in sea and lake waters, but have also produced interesting limits in
the search of cosmic neutrino sources, dark matter and exotic
particles. They are the precursors of much larger telescopes that will
be operating in the coming years.

\section*{Acknowledgements}
We gratefully acknowledge the financial support of the Spanish
Ministerio de Ciencia e Innovaci\'on (MICINN), grants
FPA2009-13983-C02-01, ACI2009-1020 and Consolider MultiDark
CSD2009-00064 and of the Generalitat Valenciana, Prometeo/2009/026.

\end{document}